\documentclass[aip,amsmath,amssymb,reprint]{revtex4-1}

\usepackage{graphicx,epsf}% Include figure files
\usepackage{dcolumn}% Align table columns on decimal point
\usepackage{bm}% bold math
\usepackage{braket}
\usepackage{siunitx}
\usepackage{xcolor}
\usepackage[normalem]{ulem}

\newcommand{\red}{\textcolor{black}}

\begin{document}

\title[]{Electromagnetic field emitted by core-shell semiconductor nanowires driven by an alternating current}

\author{Miguel Urbaneja Torres}
\affiliation{Department of Engineering, Reykjavik University,
              Menntavegur 1, IS-102 Reykjavik, Iceland}

\author{Kristjan Ottar Klausen}
\affiliation{Department of Engineering, Reykjavik University,
              Menntavegur 1, IS-102 Reykjavik, Iceland}

\author{Anna Sitek}
\affiliation{Department of Engineering, Reykjavik University,
              Menntavegur 1, IS-102 Reykjavik, Iceland}
\affiliation{Department of Theoretical Physics,
              Faculty of Fundamental Problems of Technology,
              Wroclaw University of Science and Technology,
              Wybrze{\.z}e Wyspia{\'n}skiego 27, 50-370 Wroclaw, Poland}

\author{Sigurdur I. Erlingsson}
\affiliation{Department of Engineering, Reykjavik University,
              Menntavegur 1, IS-102 Reykjavik, Iceland}

\author{Vidar Gudmundsson}
 \affiliation{Science Institute, University of Iceland, Dunhaga 3,
              IS-107 Reykjavik, Iceland}

\author{Andrei Manolescu}
%\email{manoles@ru.is}
\affiliation{Department of Engineering, Reykjavik University,
              Menntavegur 1, IS-102 Reykjavik, Iceland}

\begin{abstract}
We consider tubular nanowires with a polygonal cross-section. In this
geometry the lowest energy states are separated in two sets, one of
corner and one of side-localized states, respectively. The presence of an
external magnetic field transverse to the nanowire imposes an additional
localization mechanism, the electrons being pushed sideways relatively
to the direction of the field. This effect has important implications on
the current density, as it creates current loops induced by the Lorentz
force. We calculate numerically the electromagnetic field radiated
by hexagonal, square, and triangular nanowires. We demonstrate that,
because of the aforementioned localization properties, the radiated field
can have a complex distribution determined by the internal geometry of the nanowire. We suggest that measuring the field in the neighborhood of the nanowire could be the basic idea of a tomography of the electron distribution inside it, if a smaller receiver antenna could be placed in that zone.
\end{abstract}

\maketitle

\section{\label{sec:introduction} Introduction}

Nanowires based on semiconductor materials 
have emerged as an advantageous platform for the
realization of diverse applications in multiple areas
such as photonics \cite{Gudiksen2002,PAUZAUSKIE2006,You2019,Balaghi2019}, electronics \cite{Duan2001,Zheng2013,Porro2017,Bhuyan2018}, photovoltaics \cite{Adachi2013,Kuang_2013,Vismara2019}, or
topological Majorana physics \cite{Mourik2012,Das2012}.
With diameters of tens to hundreds of nanometers, the microscopic
crystal structure of the materials used becomes evident, resulting
in polygonal cross sections. Nanowires based on III-V semiconductors
fabricated by bottom-up growing techniques are commonly hexagonal
\cite{Blomers13,Jadczak14}, but other geometries have been obtained,
like triangular \cite{Qian04-2,Dong2009}, square \cite{Fan06} and
dodecagonal \cite{Rieger15}. Nanowires of core-shell type can also
have distinct core and shell polygonal shapes, like a triangular shell
with a hexagonal core \cite{Yuan15,Lund2019} or the other way around
\cite{Dick2010}. 

By using various semiconductor materials, growing conditions, or
controlling the thickness of the core or of the shell, the nanowire
properties can be fine-tuned for desired or potential physical behavior
\cite{Rieger12-2}.  In particular, with an insulating core surrounded
by a conductive (doped) shell, the nanowire functions as a tubular
conductor. Such a conductive nanotube may also be obtained by appropriate band
alignment  at the core-shell interface \cite{Pistol08,Wong2011,Long2012},
or via the Fermi level pinning at the outer surface of the nanowire
\cite{Heedt16}.

In the case of a polygonal shell, a group of low energy electrons,
including those in the ground-state, tend to localize at the corners of
the shell \cite{Ferrari09,Bertoni11a,Royo13,Fickenscher2013}. These corner
localization can be seen as the effect of particle accumulation at the
bending of a two-dimensional quantum wire \cite{Sprung}, representing
a transverse polygonal cross section of the shell, or at the regions
with the greatest surface curvature along the prismatic tubular shell
\cite{Bhattacharya_2016}. For a polygon with $N$ corners, and including
the electron spin, there are $2N$ states localized in the corners.
Depending on the shell thickness and sharpness of the corners, these
states are separated by an energy interval varying from a few meV to
tens of meV from the next $2N$ states, which are localized on the sides
of the polygon \cite{Sitek15}.  This corner and side states correspond to low-energy transverse modes of a prismatic shell, where, due to the longitudinal 
motion, they generate states localized along edges or 
facets, respectively.

The controlled fabrication of core-shell nanowires and their
internal structure motivated significant theoretical research
on the electronic states in tubular prismatic geometry, i.\ e.\ in the shell, relevant for optical response
\cite{Ballester12,Sitek18,Urbaneja19_2}, electrical \cite{Urbaneja18},
and thermoelectrical conductivity \cite{Erlingsson17}, or topological
physics \cite{Manolescu17,Stanescu2018,Klausen20}.  At the same time,
apart from the remarkable efforts for fabrication, relatively less
experimental research has been devoted to the electrons localized in the shell. Notably,
transport experiments showed oscillations of the longitudinal conductance in the presence of
a magnetic field, related to flux periodicity \cite{Gul14,Heedt16}. More recently, the transverse conductance and Coulomb blocking effects have been investigated in triangular nanowires \cite{Lund2019}. Also, optical emission excited by highly focused X rays indicated a nonuniform electron distribution in hexagonal multishells \cite{Criado2012}. In addition,  
the localization of electrons on facets of different thicknesses a single hexagonal shell has also been probed via photoluminescence experiments \cite{Sonner19}.  
%However, more systematic experimental evidence of the effect of the prismatic geometry of the nanowire on the electron distribution in the tubular shell, and especially the corner and side localized electronic states, with a considerably large energy gap separating them, has not been achived yet. In this paper we explore theoretically the possibility to observe quantum localization effects in the shell in the electromagnetic field radiated by the nanowire.
However, more systematic experimental evidence of the quantum localization
of electrons in the tubular prismatic geometry, and especially the presence of corner
and side localized states with an energy gap separating them, has not
been achieved yet. In this paper we explore theoretically the possibility
to relate the quantum localization in the shell to the electromagnetic
field radiated by the nanowire.

An interesting application of semiconductor based nanowires is
as elements of optoelectronic circuits, where they could function
as nanoantennas, with emitter and/or receiver function.  It was already 
demonstrated that individual nanowires made of InAs were able to detect 
electromagnetic radiation in the terahertz domain \cite{Vitiello11}. 
Later on, nanowires built from InP, have been studied by Grzela et al., 
in the near infrared domain (850 nm), and shown capable of directional 
emission and absorption via the internal Mie modes \cite{Grzela12,Grzela14}.    
More recently a multishell design of cylindrical nanowires has been proposed 
to achieve a superdirective emission in the optical domain \cite{Arslan18}. 

This interest in nanoantennas based on nanowires also motivates us to  consider the electromagnetic
fields radiated by a polygonal core-shell nanowires due to an alternating
current driven along the nanowire.  We focus on core-shell nanowires
made of semiconductors, where the charge transport occurs only within
the shell, which acts as a tubular conductor, whereas the core is an
insulator.  This situation has been obtained by doping only the material
of the shell, but not that of the core \cite{Gul14,Lund2019}.  
If the radiated field of our nanowire, seen as an emitter nanoantenna, can be explored by a separate receiver nanoantenna, then the current and charge distribution within the nanowire could be obtained by solving an inverse problem. We emphasize again that the quantum localization of the electrons in a tubular prismatic geometry is not trivial, and offers a variety of application and manipulation possibilities (including contactless \cite{Sitek17}), which have not been investigated experimentally yet.

First we compute
numerically the electronic quantum states for selected geometries, and then the current
along an infinite prismatic shell driven by a time-dependent harmonic
voltage bias.  In this work we intend to emphasize the consequences of
the internal geometry of such a nanowire, i.e.\ with a tubular prismatic
shape, and of the associated localization properties, on the radiated
electromagnetic field.  In principle such effects can be observed in
the neighborhood of the nanowire, i.e. close enough to the lateral
surface, where the anisotropic distribution of currents should lead to complex field distribution. Whereas with increasing the distance between the observation point and the nanowire the internal current distribution becomes less and less important, until eventually the radiated field looks
similar to that of a simple, unstructured wire of finite length.
For these reasons we shall consider nanowires of an infinite length, and
we shall calculate the radiated field in the proximity of the nanowire.

We describe our numerical methodology in Section \ref{sec_model}. Then, 
in Section \ref{sec_results} we
qualitatively analyze the signature of the
prismatic tubular geometry of the nanowire on the structure of the
radiated electromagnetic field.  We discuss the implications of the prism
edge and facet localization, and in particular the anisotropy of the
radiated field.  Furthermore, we explore the variation of the radiated
field when applying an external magnetic field perpendicular to the nanowire.
The anisotropy of the radiated field is limited to the distances much
smaller than the nanowire length, but depending on the geometry of the
nanowire and on other parameters, that zone can be considerably large
compared to the nanowire radius. Finally, in Section \ref{sec_conclusions} we collect our conclusions and comments on the tomography idea.

\section{\label{sec_model} Model and methods}

We start by considering a system of non-interacting electrons confined in
a polygonal ring. The model begins with
a circular disk situated in the plane $(x,y)$ which is discretized in
polar coordinates \cite{Daday11}. On this grid, we superimpose polygonal
constraints and we retain the points which lie inside the resulting
boundaries, excluding the rest. With this method we define hexagonal,
square, and triangular cross-sections. Further, we consider the
nanowires to be infinite, assuming free particle propagation along their
length, i.e., in the $z$ direction. The Hamiltonian of the system is
then expressed as follows:
\begin{equation}
\label{ham}
 H = \frac{(-i\hbar \nabla + e\bm{\mathrm{A}})^2}{2m_{\rm eff}} 
-g_{\rm eff}\ \mu_{\rm B}\ {\bm \sigma}{\bf B} \ ,
\end{equation}
where ${\bf B}=(B_x,B_y,0)$ is an external magnetic field, transversal to the nanowire length, and
$\bm{\mathrm{A}}$ is the corresponding vector potential. The vector ${\bm r}=(x,y,z)$ defines the positions inside the shell, $e$ is the electron charge, $m_{\rm eff}$ and $g_{\rm eff}$ are the effective electron mass and bulk g-factor of the material considered, $\mu_{\rm B}$ is the Bohr's magneton and  ${\bf \bm{\sigma}}=(\sigma_x,\sigma_y,\sigma_z)$ are the spin Pauli matrices.

The Hilbert space associated with the polar grid is spanned by the
position vectors $\ket{q} = \ket{r_q,\phi_q}$, where $r_q$ and $\phi_q$ are
the radial and angular coordinates of site $q$, respectively, from which the spin
is excluded. In order to obtain the eigenstates of the Hamiltonian (\ref{ham}),
we solve the problem in two steps: we first obtain the transverse
eigenstates $\ket{a} = \sum_q \psi(q,a) \ket{q}$ and eigenvalues $E_a$
($a$ = 1,2,3...), for ${\bf B} = 0$, i.e.\ the states of an electron in the 
polygonal ring representing the cross section of the nanowire. Then,
we retain the transverse states with lowest energies, 
%(up to 4$N$), and
%the first $2N$ modes (where $N$ is the number of corners or sides of the polygon), 
and together with the plane wave vectors in the $z$
direction, $\ket{k} = \mathrm{exp}(ikz)/\sqrt{L}$, where $L$ is the
length of the nanowire (considered infinite in our model), and with the spin
states $\ket{s} = \pm 1$, we form the basis $\ket{aks}$. Finally, for
${\bf B}\neq 0$, we diagonalize numerically the total Hamiltonian (\ref{ham}) in this basis,
using a discretized series of $k$ values, and obtain the eigenvalues
$E_{mks}$ ($m=1,2,3,...$) and eigenstates $\ket{mks}$ of the Hamiltonian (\ref{ham}),
expanded in the basis $\ket{aks}$. 
The number of transverse states $\ket{a}$ needed to reach numerical convergence was typically $4N$ (excluding spin). The convergence was checked in several cases for an even larger basis set, and no change of the final results was observed.
The numerical diagonalizations were performed with the Lapack library.

With this approach we obtain the first and the second lowest-energy groups of states
for the infinite prismatic tubular nanowire, which are localized along the edges or along 
the facets of the prism, respectively. The charge density associated to these states is
\begin{equation}
\rho(\bm{r}) = - e \sum_{mks} \! {\cal F} \left(\frac{E_{mks}-\mu}{k_{B}T}\right) 
\left(|\braket{\bm{r}|mks}|^2 - n_d\right)\! \ ,
\end{equation}
where $-e$ is the electron charge and $n_d$ represents the density of the ionized donors.
The next step is to compute the current 
density ${\bm J}$ inside the shell, carried by the edge- and facet-localized states:
\begin{equation}
\label{expected_current}
\bm{J}(\bm{r}) = \sum_{mks} \! {\cal F} \left(\frac{E_{mks}-\mu}{k_{B}T}\right) 
\braket{mks|\bm{j}(\bm{r}-\bm{r}_0)|mks}\! \ ,
\end{equation}
with ${\cal F}(u) = 1/[\mathrm{exp}(u)+1]$ being the Fermi function
and $u =(E_{mks}-\mu)/k_{B}T$, where $\mu$ is the chemical potential,
$T$ the temperature, and $k_B$ the Boltzmann's constant. The operator
$\bm{j}(\bm{r},\bm{r}_{0}) = e[\delta(\bm{r}-\bm{r}_0)\bm{v} +
\bm{v}\delta(\bm{r}-\bm{r}_0)]/2$ describes the contribution to the total
current density at the position $\bm{r}$ from an electron situated at
$\bm{r}_{0}$ and moving with a velocity $\bm{v}$. The velocity is defined
by the operator $\bm{v} = i[H,\bm{r}_0]/\hbar$ \cite{Messiah}.

If no longitudinal voltage bias is applied, the system is in equilibrium
and the contributions to the total current of electrons moving with
opposite velocities in the $z$ direction compensate each other. When a
voltage bias is considered, these contributions no longer cancel out and
the nanowire carries a non-zero total current. In order to simulate the
effect of a voltage bias, we create in our system an imbalance between
electrons in states corresponding to positive and negative velocities,
i.e., $\partial E_{mks}/ \partial k > 0$, and $\partial E_{mks}/ \partial
k<0$, respectively \cite{Datta}. Thus, we consider two different values,
harmonically time-dependent, for the chemical potentials $\mu_{+}$ and
$\mu_{-}$ for carriers moving in opposite directions, with $\mu_{\pm}
= \mu \pm V \sin(\omega t)$, where $\omega$ is the frequency, $2V$ the
bias amplitude and $\mu$ is the static chemical potential, which is
determined by the carrier density at equilibrium.

Once the current density distribution is obtained, we calculate the 
time dependent scalar and vector potentials outside the nanowire,
\begin{equation}
\label{scalpot}
V(\bm{r},t) = \frac{1}{4 \pi\varepsilon_0} \int \frac{\rho (\bm{r'},t)}{|\bm{r}-\bm{r'}|} \mathrm{d}{\bm r'} \ ,
\end{equation}
\begin{equation}
\label{vecpot}
\bm{A}(\bm{r},t) = \frac{\mu _0}{4 \pi} \int \frac{\bm{J} (\bm{r'},t)}{|\bm{r}-\bm{r'}|} \mathrm{d}{\bm r'} \ , 
\end{equation}
where $\varepsilon_0$ and $\mu_0$ are the vacuum electric permittivity and magnetic permeability, respectively, and the integration is carried out inside the shell. 

To properly obtain the vector potential we consider the approximation $r \ll L$.
Having the vector potential it is straightforward to obtain the electromagnetic radiated field using the relations

\begin{equation}
\label{field}
\bm{E} = - \nabla V - \frac{\partial \bm{A}}{\partial t},\hspace{0.25cm} \mathrm{and} \hspace{0.25cm}\bm{B} = \nabla \times \bm{A} \ .
\end{equation}
We thus neglect the retardation effects and displacement currents, since we will consider relatively low frequencies and quasi-stationary currents
%and $\nabla {\bf A}=0$ 
\cite{Notaros}.  In all our calculations the contribution from the scalar potential was also very small. 
%\red{In fact, in Eq.~(\ref{scalpot}) we neglected the polarization effects inside the nanowire, both due to the lattice dielectric response and due to the electron-electron Coulomb interactions.  In principle these screening corrections reduce the scalar potential outside the nanowire, at least for low electron densities and low frequencies that we are considering in our present work. Therefore a more elaborated model replacing Eq.~(\ref{scalpot}) will not significantly change our results.} 
\red{In fact, in Eq.(4) we neglected the polarization effects inside the nanowire, both due to the lattice dielectric response and due to the electron-electron Coulomb interactions.  In principle such screening corrections lead to a dielectric constant $\varepsilon > \varepsilon_0$ inside the nanowire and reduce the scalar potential everywhere,  for the electron  densities and frequencies considered  in  our  work.   Hence, with a more elaborated model replacing Eq.(4) the contribution of $V$ to our results will remain negligible.}

The next step is to estimate the frequency domain we can consider with our method. 
Typical nanowires made of semiconductor materials such as InAs, InP
or GaAS have electron mobilities $\mu_e$ in the range of 400-6000
$\mathrm{cm^2/(Vs)}$, with scattering times in the range of tens of femtoseconds \cite{Timm13,Joyce2016}. Thus, for frequencies in the RF domain, from kHz to GHz, achievable within an electric
circuit connected to the nanowire, we can reduce the electron damping
and losses in the semiconductor nanowire to the static resistivity,
and ignore the dynamic corrections. For higher frequencies, towards the optical domain, the effects of the anisotropic and dynamic permittivity must be taken into account both inside and outside the nanowire, for example like in Ref. \onlinecite{Urbaneja19_2}.

In our present RF regime the current can be injected into the nanowire through source-drain contacts
covered with top-gate electrodes. This
method has been used in a recent experiment to study the current output in
core/shell GaSb/InAsSb NWs as a function of AC gate voltage and frequency,
using a four-probe method \cite{Ganjipour_2014}. In our work we assume
the contacts are far from each other, at several microns distance,
possibly at the two ends of the nanowire, such that the length of the
resulting antenna is much larger than the radius.  As we mentioned in the
Introduction, we are interested in the field distribution in the vicinity
of the nanowire resulting from the prismatic geometry.  At large enough
distances from the nanowire its internal prismatic geometry becomes
irrelevant, but instead, the geometry of the contacts may play a role.
In addition, in practice the contacts contribute to the impedance of the
nanowire, and therefor to the current strength, and also possibly to
the current distribution in the contact region.  Nevertheless, including
these details in our model is beyond the scope of our present work, where
we want to concentrate on the primary implications of current carrying
tubular prismatic shell.

We consider InAs bulk parameters for the nanowire: $m_{\rm eff}=0.023m_e$
and $g_{\rm eff}=-14.9$. We use a voltage bias corresponding to $V =
2.5$ meV, and we consider a test frequency of 1 MHz.  We are mostly
interested in the angular distribution of the radiated field around
the nanowire, and for this purpose the frequency plays no qualitative
role. It quantitatively affects though the conductivity, which drops down
in the GHz domain due to damping, but also the power density, which - in
principle - increases proportionally to the frequency squared.  For our purpose we
evaluate the power density of the radiated field with the time-averaged magnitude of the
Poynting vector:

\begin{equation}
\label{Poynt}
\bm{\overline{S}(\bm{r})} = \frac{1}{T} \int_0^T \frac{1}{\mu_0} (\bm{E}(\bm{r},t) \times \bm{B}(\bm{r},t))\hspace{1mm}dt \ .
\end{equation}

In all the examples shown in this article the external radius of the
shell (measured from the center of the polygon to one corner) and the
thickness of the facets are $R_{\rm ext}=30$ and $t=6$ nm, respectively.

\section{\label{sec_results} Results}

For sufficiently thin shells with polygonal cross section, like ours, the electrons with the
lowest energies are localized in the corners of the polygon and the
electrons in the next layer of energy states are localized on the sides
\cite{Sitek15}. The corner and side states are energetically separated by
a gap interval that depends on the  shape and on the aspect ratio of the
polygon, i.\ e. the ratio between the thickness and external radius of the cross section. It increases with decreasing the shell thickness or the number
of corners, and hence, in such a structure the subspace of corner states  
is potentially robust to many types of perturbations \cite{Sitek19}.

\subsection{No external magnetic field}

% -----------------------------------------------------------------
\begin{figure}[t]
\centering
\includegraphics[scale=0.44]{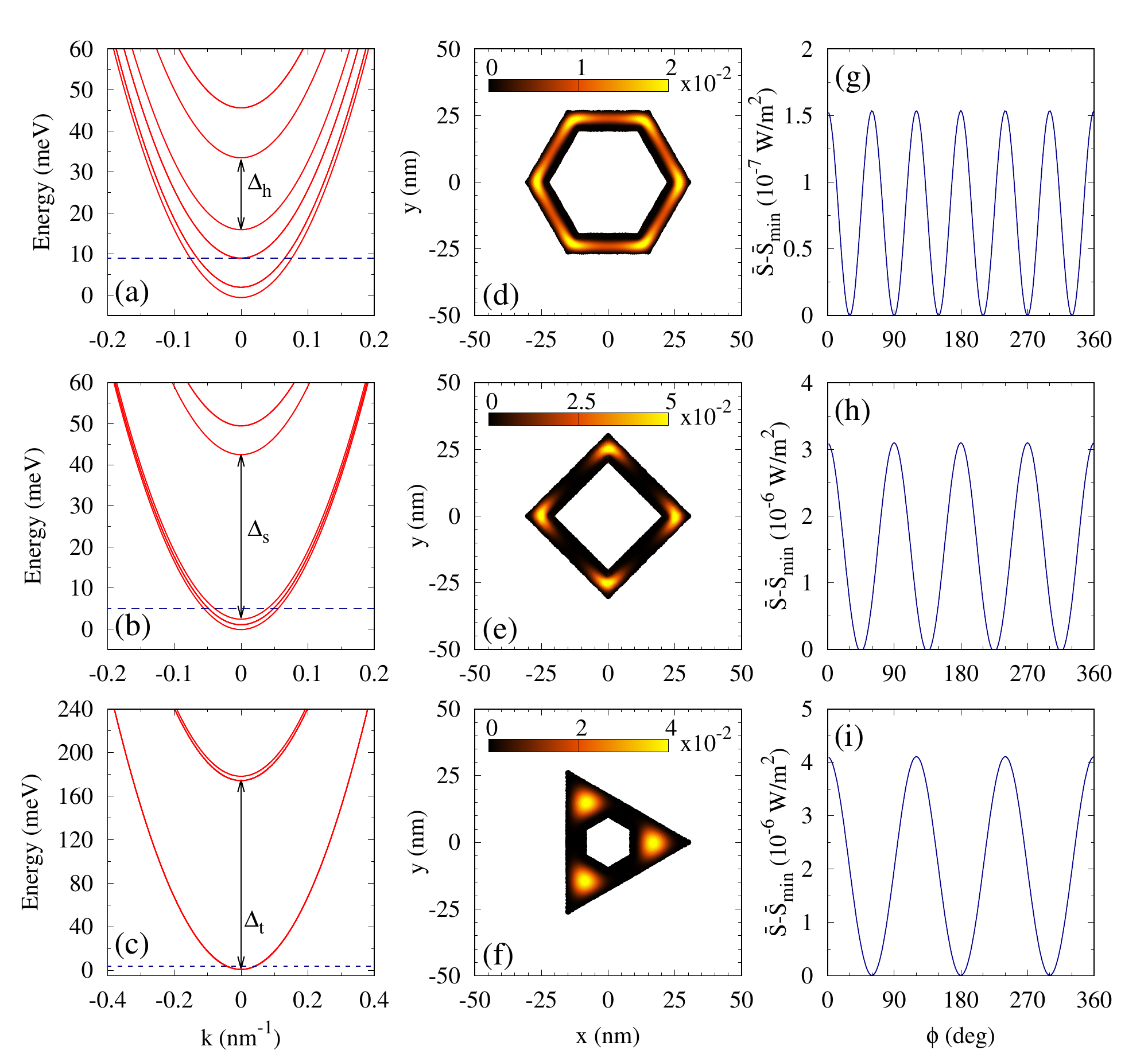}
\caption{(a-c): Energy spectra obtained for the three polygonal
shapes. The blue-dashed line marks the chemical potential $\mu$ in
equilibrium for a carrier density of 
%$n = 10^{-4}$ nm$^{-3}$, 
$n = 10^{17}$ cm$^{-3}$,
the arrows indicate the energy gap between the groups of corner and side
states. (d-f): Time derivative of the current density ($d\bm{J}/dt$)
inside the nanowire when its amplitude is maximal, i.e, at an instant in
time when  $\mu_{\pm} = \mu$. The color scale units are $\mathrm{A/(s \ nm^2)}$. (g-i): relative difference of the time-averaged magnitude of the Poynting vector
(radiated power density) as a function of the angle at a distance of
$2R_{\mathrm ext}$ from the center of the nanowire. The
initial $\ang{0}$ is set at the rightmost corner of the shell in the
figures and increases counter-clockwise.}
\label{First}
\end{figure}
% -----------------------------------------------------------------

In Figs.~\ref{First}(a-c) we show the energy spectra of nanowires with
hexagonal, square, and triangular-hexagonal cross sections assuming 
no external field. The blue dashed horizontal lines indicate the chemical potential level. The energy  intervals between
corner and side states are $\Delta_h$ = 21.2 meV for the hexagon,
$\Delta_s$ = 34.4 meV for the square and $\Delta_t$ = 171.1 meV for
the triangle. The chemical potential (blue-dashed line) corresponds to an electron density of 
%$n = 10^{-4}$ nm$^{-3}$, 
$n = 10^{17}$ cm$^{-3}$,
a regime which can be achieved experimentally \cite{Heedt16}. 
%\blue{\sout{The triangular shell with a hexagonal core has been %obtained
%by the authors of Ref.\ \cite{Lund2019}.}}
For this carrier concentration the plasma frequency is of the order of $10^{13}$ Hz, far enough from our frequency domain, such that our quasi-static description of the electromagnetic field is reliable.
Because of the spin and rotational symmetries of these
geometries the states can be two- or four-fold degenerate, such that
multiple energy levels overlap in each spectra. Each group of corner
and side states consists of twelve levels for the hexagon, eight for
the square, and six for the triangle, and the corresponding degeneracy
patterns are 2442/2442, 242/242, and 24/42, respectively \cite{Sitek16}.

When the time dependent harmonic voltage is applied to the nanowire, the
total current is zero at those instants of times when $\mu_{\pm} = \mu$,
i.e., when $\sin(\omega t) = 0$. However, at these instants the amplitude
of the radiated electric field is maximal, as it depends on the time derivative
of the current density, through the vector potential calculated
with Eq.~(\ref{vecpot}). On the contrary,
when $\sin(\omega t) = 1$, and the applied voltage bias is maximum,
i.e., $\mu_{+}-\mu_{-} = 2V$, the current reaches
its maximum, but its time derivative vanishes, and so does the radiated
electric field. It is important to note that, when $\mu_{\pm}
\ne \mu$ or $\sin(\omega t) \ne 0$, depending on the amplitude of the
harmonic voltage applied, electrons may move between different energy
levels, which translates into changes of localization, and hence of the current
distribution and radiated field, within one period of time.

In Figs.~\ref{First}(d-f) we plot the time derivative of the current
density, $dJ/dt$, inside the nanowire cross-section. 
For simplicity we only show it in the case of $\mu_{\pm} =
\mu$, when the radiated field intensity is maximal.   
We use a carrier density sufficiently low in order to obtain a chemical
potential at equilibrium ($\mu$) within the low energy bands. 

In the absence of an external magnetic field the distributions
shown for $dJ/dt$ are qualitatively similar to the electron 
concentrations and current density at equilibrium \cite{Urbaneja18,Urbaneja18b}. Obviously, the imposed voltage bias creates an imbalance between carriers moving in opposite directions along the length of the nanowire. In our case the total current driven along the nanowire ranges between a few nA to a few $\mathrm{\mu A}$ for voltage
bias amplitudes of 1-10 meV \cite{Urbaneja18}.

In Figs.~\ref{First}(g-i) we show the angular variation of
the power density of the radiated electromagnetic field at the
distance $d=2R_{\mathrm ext}$ from the center of the nanowires. The
reference minimal values are $\overline{S}_{min} = 3.01,\hspace{1mm}
3.06,\hspace{1mm} 1.76 \times 10^{-4}\hspace{1mm}\mathrm{W/m^2}$ for the
hexagonal, square and triangular nanowires, respectively. As expected
from the current density distribution, the resulting field captures
the internal geometry of the nanowires manifested by a number of peaks
which is equal to the number of corners. The electrons localized in each
corner form quasi-independent current channels which can be considered
as individual nanowires.

In classical terms, one can attribute to the radiated field a singular behavior due to the quasi one or two dimensional current distributions within the prismatic shell with sharp corners and thin sides \cite{Bladel}.  However, here we take into account properly the transverse degree of freedom, consistently with the quantum localization of the electrons, along parallel current filaments with transverse geometry and different strengths. This special current distribution
leads to multipole variations in the radiated field, of which the lower one will have the largest strength. We thus expect weakly imposed quadrupole, or hexapole, or higher order pattern in the radiation field, or more unusual patterns, as we will see in Subsection \ref{magfield}, due to the presence of a magnetic field.

% -----------------------------------------------------------------
\begin{figure}[t]
\centering
\includegraphics[scale=0.65]{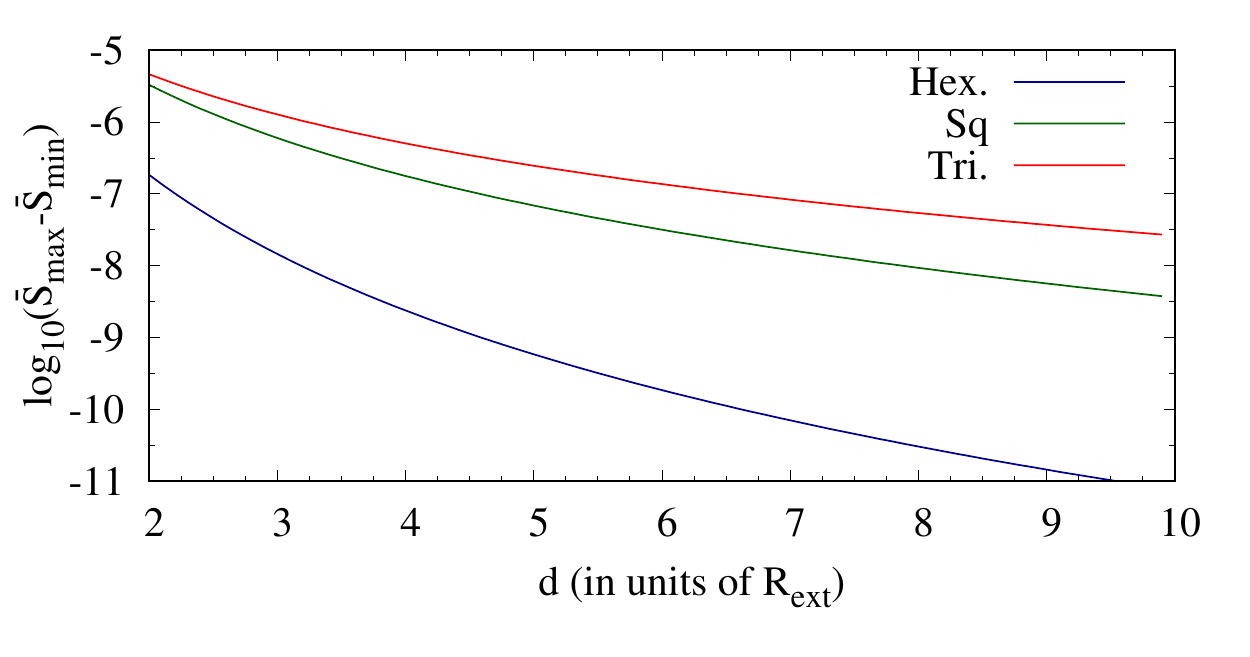}
\vspace{-4mm}
\caption{Decay of the relative difference, in log-scale, between minimum and maximum values of the radiated power density as a function of the distance from the center of the nanowire.}
\vspace{-2mm}
\label{Decay}
\end{figure}
% -----------------------------------------------------------------

Before that, in Fig.~\ref{Decay} we show how the differences between the
maximal and the minimal value of the power density curves, shown
in Fig.~\ref{First}(g-i), decay with the distance from the center of the 
nanowire, within the near
field zone, between ${d = 2R_{\mathrm ext}}$ and ${d = 10R_{\mathrm
ext}}$. The fastest decay is observed for the hexagonal nanowires which has
the least anisotropic geometry, closest to circular. As expected, the
decay is less pronounced with decreasing the number of corners of the
nanowire. The signals originating from the square and the triangular
nanowires have similar intensity at the position ${d = 2R_{\mathrm
ext}}$, but they decay much slower around the triangular than
around the square nanowire. The difference at ${d= 10 R_{\mathrm ext}}$
is around  $10^5$, $10^3$ and  $10^2$ times smaller for the hexagonal,
square, and triangular nanowires, respectively.

\subsection{\label{magfield} Effects of a transverse external magnetic field}

In this subsection we consider a uniform magnetic field perpendicular to the 
nanowire axis. Such a 
magnetic field may drastically perturb the current
distribution within the tubular shell, mainly because the field component
normal to the nanowire surface, which is responsible for the Lorentz
force, varies with the angular coordinate.  In the case of a circular
cross section, and for strong enough fields, the electrons situated
on the sides of the tubular shell, relatively to the direction of the
magnetic field, experience nearly zero local magnetic field, and move on
snaking trajectories along the wire, in opposite directions.
On the contrary, the electrons situated on the top or bottom regions tend to
perform local cyclotron loops
\cite{Tserkovnyak06,Ferrari08,Manolescu2013,Rosdahl14,Chang16}.  

\begin{figure}[t]
\centering
\includegraphics[scale=0.60]{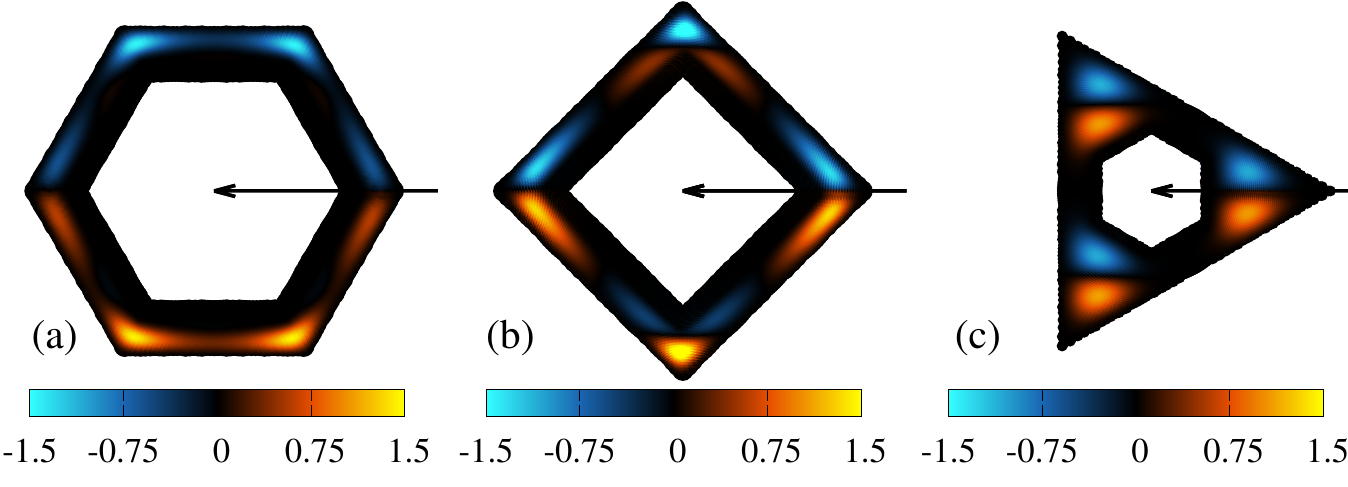}
\caption{Current distribution in equilibrium inside the polygonal shells for 
%$n = 10^{-4}$ $\mathrm{nm^{-3}}$ 
$n = 10^{17}$ cm$^{-3}$
when a magnetic field of $B = 1$ T is applied perpendicular to one of the edges for the hexagonal (a), square (b) and triangular (c) shells. The units of the color scale are $\mathrm{nA/nm^{-2}}$.}
\label{jDensity}
\end{figure}

The situation becomes more complex for a polygonal shell, where the
localization resulting from the geometry competes with the one induced by
the magnetic field \cite{Ferrari09,Urbaneja18}.  
As we pointed out previously, when no longitudinal voltage is applied
over the nanowire, the total current vanishes due to equal number of
electrons moving to the left and right.  This holds for both zero and
non-zero transverse magnetic field.  In the case of the current density
$\bm{J}$, which reflects the underlying charge and velocity distributions,
the presence of a non-zero magnetic field leads to splitting due to
the Lorentz-force.  This can be seen in Fig.~\ref{jDensity}, where
the equilibrium current density is split into channels consisting of
electrons with positive and negative velocities, the former ones pushed
to the right side of the sample, while the latter to the left side. The
current density takes the form of loops along the $z$ axis of the nanowire
which close up at the ends of it (i.\ e. at infinity in our model). Then,
even if the total integrated current is still zero, the current density
is not compensated locally.

% -----------------------------------------------------------------
% -----------------------------------------------------------------

However, in equilibrium, the channels that form each loop are compensated,
i.e., the same current flows in both directions, with each channel
paired with the one on the opposite side of the geometric symmetry axis
relative to the magnetic field direction. Depending on the localization
strength, the pairing can happen within the same corner or side or
on opposite ones. Thus, we observe that in the case of the hexagon
(Fig.~\ref{jDensity}(a)), for which the localization is the weakest, there
is one main loop formed on the two sides parallel to the magnetic field
direction, where the snaking states are formed, and two additional loops
originating from the split maxima in the corners situated at angles
$\ang{0}$ and $\ang{180}$ with respect to the  magnetic field. In the case
of the square, Fig.~\ref{jDensity}(b), the situation is similar, with
one main loop that captures most of the current and two additional ones
on the corners perpendicular to the field direction. For the triangular
case, Fig.~\ref{jDensity}(c), where the localization is the strongest,
there are three loops and a pair of opposite channels in each corner.

\begin{figure}[t]
\centering
\includegraphics[scale=0.44]{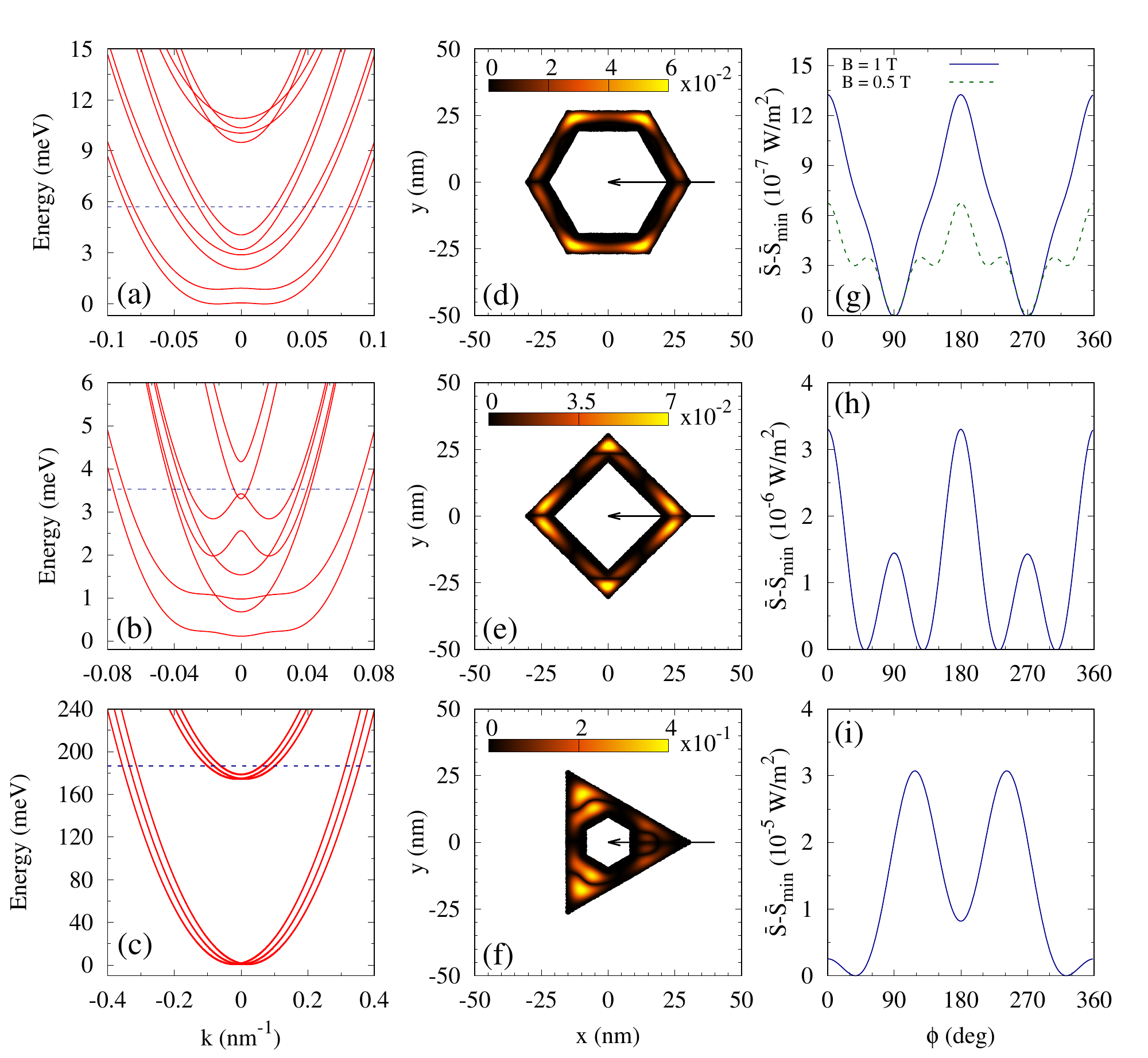}
\caption{As in Fig.~\ref{First}, the energy spectra (a-c), but now in the presence of an external magnetic field of $B = 1$ T perpendicular to one of the edges of the nanowire indicated by the arrows in (d-f). The power distribution outside the nanowire no longer reproduce the geometry of the cross section (g-i). The additional dashed green curve in (g) corresponds to the case of $B = 0.5$ T, shown in order to illustrate the evolution of the curve. The chemical potentials correspond to a carrier density of 
%$n = 10^{-4}$ nm$^{-3}$
$n = 10^{17}$ cm$^{-3}$
in panels (a) and (b) while for the triangular nanowire in (c)   
%$n = 10^{-3}$ nm$^{-3}$.
$n = 10^{18}$ cm$^{-3}$. 
The color scale units are $\mathrm{A/(s\ nm^2)}$.}
\label{Second}
\end{figure}

If the magnetic field is strong enough, it can mix the two groups of corner and side states, leading to complex changes in the energy spectra and in the current distributions. The dispersion with respect to the wave vector $k$ when a magnetic field is applied perpendicular to one of the edges is shown in Figs.~\ref{Second}(a-c). We use the same carrier density 
%($n = 10^{-4}$ $\mathrm{nm^{-3}}$) 
$n = 10^{17}$ cm$^{-3}$
for the hexagon and the square cases, so that the position of the chemical potential would be at the level of the corner states if there was no magnetic field applied, as it was shown in Figs.~\ref{First}(a-c), whereas for the triangular case we now use 
%$n = 10^{-3}$ $\mathrm{nm^{-3}}$
$n = 10^{18}$ cm$^{-3}$
in order to reach the side states.

The time derivative of the current density ($d\bm{J}/dt$), shown 
in Figs.~\ref{Second}(d-f), contrary to the current distribution, has
only either positive or negative sign, as the electrons are oppositely
accelerated or slowed down by the voltage bias, depending on its
sign. For clarity, the time-derivative of the current distribution
over the whole time period is shown in Fig.\ \ref{Supplementary_a} of the 
Appendix.

% -----------------------------------------------------------------
% -----------------------------------------------------------------

In Figs.~\ref{Second}(g-i) we show how the magnetic field affects the relative differences of the power density. The minimum values of the power density are now $\overline{S}_{min} = 1.74,\hspace{1mm} 2.23,\hspace{1mm} 15.45 \times 10^{-4}\hspace{1mm}\mathrm{W/m^2}$ for the hexagonal, square and triangular nanowires, respectively. For the hexagonal nanowire we observe a clear difference with respect to the case without external magnetic field. Instead of the six corner-related peaks we now observe only two main ones with some shoulders. The reason is that now the Lorentz force pushes the current density to the sides where the snaking states are formed, which carry most of the current. 

We note, however, that the time-average Poynting vector is an integration
over a complete time period. When the chemical potentials are imbalanced,
the nanowire quickly carries net current only in one direction and thus,
over half of a period the contribution of one of the two sides is much
smaller. On the contrary, the contribution of the two other regions where
the additional loops are formed remains relevant during the complete
period. Thus, on average, the two peaks of the maximum radiated power
density appear on the angles corresponding to the two corners pierced by
the field. For clarity, to illustrate this effect, we also show the current
density distribution over a complete period in Fig.\ \ref{Supplementary_b} of 
the Appendix. 

% -----------------------------------------------------------------
\begin{figure}[t]
\centering
\includegraphics[scale=0.44]{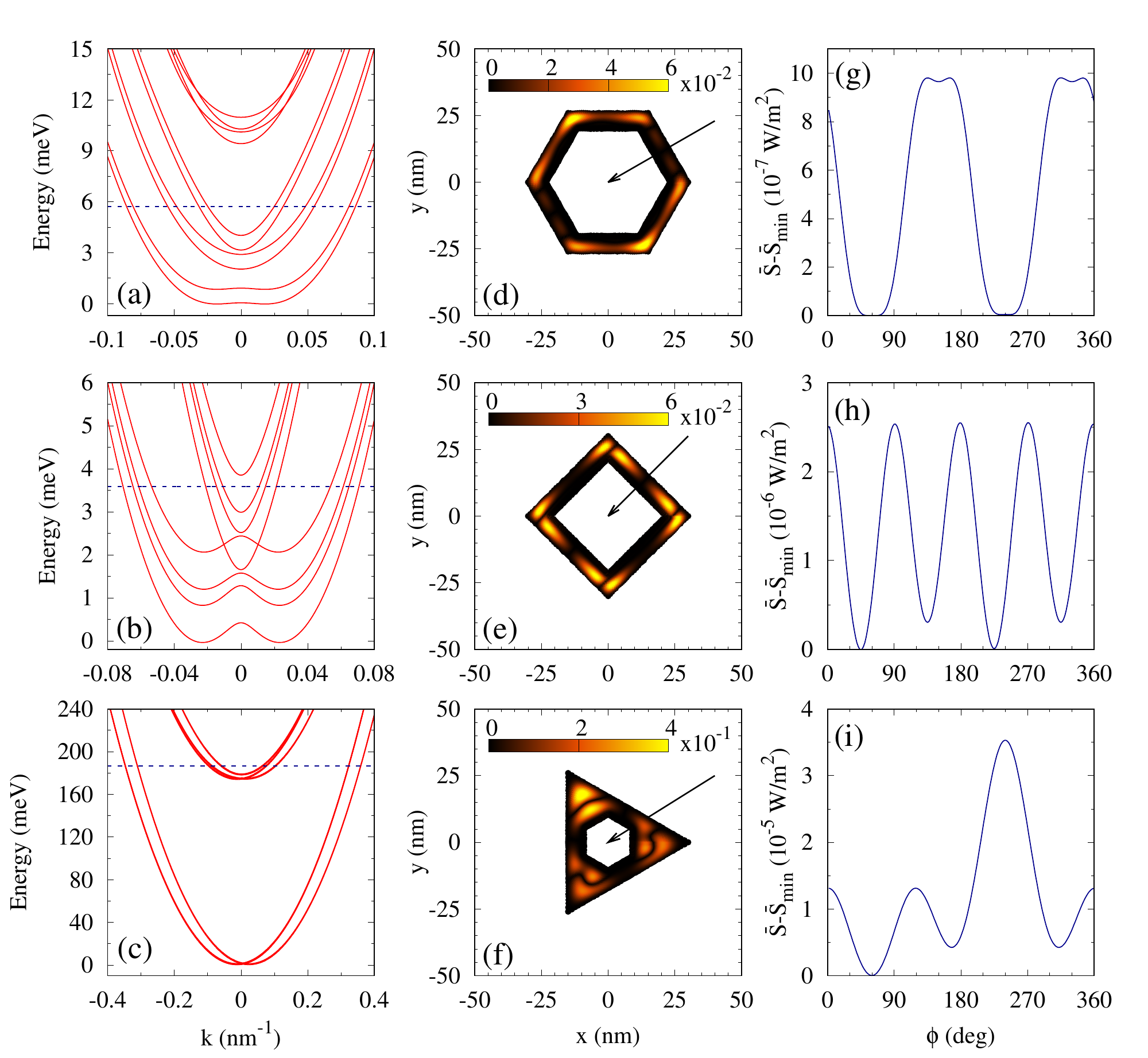}
\caption{As in Figs.~\ref{First} and \ref{Second}, panels (a-c) show the energy spectra, (d-f) the time derivative of the current density, and (g-i) the power density around the nanowire, now with the external magnetic field perpendicular to one of the facets of the nanowires in (d) and (e), and parallel to one side for the triangular case (f). The color scale units are $\mathrm{A/(s\ nm^2)}$}
\label{Third}
\end{figure}
% -----------------------------------------------------------------

The situation is, to some extent, similar in the case of the square, but
the difference in this case lies in the fact that each corner hosts current
in both directions. Thus, integrating over a period, we still observe
four peaks in the power density but now with different intensities. 

In the triangular case the corner localization is stronger
and the current distribution is locked in each corner if the carrier
concentration is too low. With 
%$n=10^{-4} \ {\rm nm}^{-3}$ 
$n = 10^{17}$ cm$^{-3}$
as before, 
the effects of a magnetic field of the order of 1 Tesla are negligible, and the
radiated field profile is similar to the one in Fig.~\ref{First}(i). For this
reason we increase the carrier density to 
%$n = 10^{-3}\ {\rm nm}^{-3}$, 
$n = 10^{18}$ cm$^{-3}$,
so that the
chemical potential crosses the next group of states, which are side localized,
for which the localization is weaker. The Lorentz force then spreads the 
current loops laterally with respect to the magnetic field.  For the
field orientation shown they are symmetrically enhanced in two corners.  
This pattern is reflected
in the power density curve where, we now observe only two clear peaks at angles
corresponding to those corners, i.e., $\ang{120}$ and $\ang{240}$.

Taking advantage of the localization mechanism induced by the external
magnetic field, it is possible to tune the directivity of the
nanowires. In Fig.~\ref{Third} we consider the same parameters as in 
the previous examples, but we change the direction of the magnetic field. 
As can be seen in Figs.~\ref{Third}, a change in the direction of the magnetic field does not only shift the power density curves, as it happens for the hexagonal case, but also creates a different localization which leads to different current density distributions. Thus, this change has also implications on the power density curves, i.e., it mainly affects the overlapping of the contributions between the main loop and the weaker ones for the hexagon and the relative difference in their amplitude for the square. 
For the triangular nanowire the rotation of the field, now parallel
to one of the sides, leads to the accumulation of current in
the corner opposite to its direction, which translates in the
formation of a dominant peak in the power density curve as seen in
Fig.~\ref{Third}(i). 
The minimal values of the power density are in this later 
case $\overline{S}_{min} = 1.74,\hspace{1mm} 2.37,\hspace{1mm}
15.62 \times 10^{-4}\hspace{1mm}\mathrm{W/m^2}$ for the hexagonal, square
and triangular nanowires, respectively. Additionally, as can be seen in
Fig.~\ref{Third}(c), because of the absence of an inversion center for
the triangle, the energy curves are no longer symmetric with respect to
the wave vector $k$ if the magnetic field points parallel to one of the sides.

% -----------------------------------------------------------------
\begin{figure}[t]
\vspace{3mm}
\centering
\includegraphics[scale=0.46]{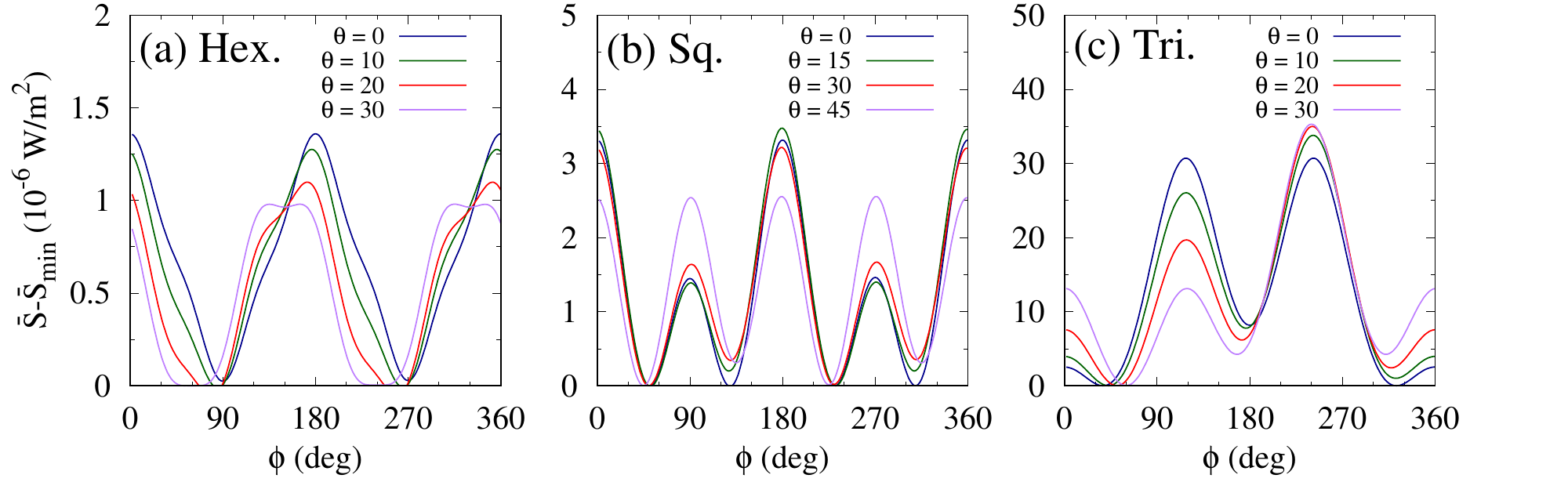}
\caption{Relative difference of the time-averaged Poynting vector as a function of the angle for different orientations of the magnetic field for the three different cross-sections: hexagonal (a), square (b), and triangular (c).  The angle of the magnetic field is varied from being perpendicular to one of the corners to being perpendicular to one of the sides in steps of \ang{10} for the hexagonal and triangular nanowires and steps of \ang{15} for the square. As in the previous cases the power density is measured at a distance of $2{R_{\rm ext}}$ from the center of the nanowire.}
\label{Final}
\end{figure}
% -----------------------------------------------------------------

Finally, for comparison, in Fig.~\ref{Final} we show the relative
difference of the radiated power density for different orientations
of the external magnetic field. For a hexagonal nanowire, where the geometric localization is the weakest, a change in the magnetic field orientation
results in an angular shift of the power density curve. 
The change of shape of the power distribution can be attributed to the 
polygonal geometry, as the electrons
are being pushed to different areas of the shell and different loops
are formed (Fig.~\ref{Final}(a)). For the square nanowire, instead,
the change of orientation of the field does not shift the curves, as
the geometry-induced localization is stronger than for the hexagonal
case, but leads to a change of the contribution from each corner
(Fig.~\ref{Final}(b)). For the triangular nanowire, for which we used
a higher carrier density 
%($n = 10^{-3}$ nm$^{-3}$), 
($n = 10^{18}$ cm$^{-3}$),
a rotation of the
magnetic field leads to a gradual decrease of the power density emitted
by the current hosted by the corner perpendicular to its direction,
as the Lorentz force pushes the current to the opposite side, where the
main loop is formed (Fig.~\ref{Final}(c)).

\section{\label{sec_conclusions} Conclusions}

Polygonal semiconductor core-shell nanowires exposed to a transverse
external magnetic field incorporate complex electron localization mechanisms
yielding a rich phenomenology. The direction and magnitude of the
magnetic field allows for the tunability of the electron localization
and creates current loops, or channels of current where electrons
travel in opposite directions. The electromagnetic field emitted by these
nanowires subjected to an alternate current along their length depends
on the underlying electron and current distribution inside the nanowire.

In order to study (some) consequences of the electron localization inside
a prismatic core-shell nanowire, we obtained the radiated power density as a function
of the angle, in the neighborhood of it. We should
mention that we on purpose avoided the term ``near field'', which is
more familiar to the electrical engineers, and implies a comparison
between the wavelength of the radiated field and the distance from the antenna.  Still, our nanowires should be considered electrically small antennas, i.\ e.\ with length much smaller than the wavelength.  However,
for our goal, the frequency, and thus the wavelength, are not essential
parameters, the relevant domain for the field anisotropy beeing given by the geometry of the prismatic shell.  

In a recent work a nonuniform optical field emitted by hexagonal
core-shell structures with imperfect internal geometry, via
space resolved photoluminescence, has been detected and associated with electron localization
within the shell \cite{Sonner19}.  In the present
paper we considered instead the radio frequency domain.  We showed that
the radiated electromagnetic field can capture the electron and current
density localization and we studied how it is affected by a magnetic
field transverse to the nanowire, with different angular orientations.

The magnitude of the power emitted by the nanowires highly depends on
the parameters used in the model, i.e., carrier density, amplitude of
the voltage bias and, especially, frequency, and it can vary by orders
of magnitudes. Rather than an in-depth study of the magnitude of the
emitted power in different situations, our goal has been to demonstrate,
more qualitatively than quantitatively, that the resulting field follows the anisotropy of the charge and current distribution within the shell of the nanowire, and has a tunable directivity, first via the geometry and internal structure of the nanowire, and second with an external magnetic field. Although we restricted our study to individual nanowires, the extension of the ideas to arrays of parallel nanowires, as nanowires are often grown, to achieve a combined effect and increased
power, is straightforward. 

If the electromagnetic field radiated by our core-shell nanowire could be measured with a receiver nanoantenna, at a sufficiently short distance, the data could be used for a tomography of the charge distribution in the nanowire.  The receiver may be another nanowire, shorter than the emitter, made either of a semiconductor \cite{Vitiello11} or a metal \cite{Zijlstra2012}.  Other solution could possibly follow the examples of the near-field optical spectroscopy of a Yagi-Uda nanoantenna \cite{Dorfmuller2011}, or using a nanoantenna array as receiver \cite{Dregely2014}. Another interesting example is a tandem of receiving-transmitting nanoantennas situated 500 nm apart \cite{Ginzburg2011}.  
Indeed, technical challenges are inevitable, like approaching the two emitter and the receiver at a distance within the emitter near field pattern, adjusting the relative distance and angular position of the two elements, etc. For example like in the atomic force or scanning tunneling microscopy. 
In addition, depending on the specific structure of the receiver, an interaction with the emitter nanowire is possible, and that should be included in the reconstruction of the current distribution.
The design of a proper setup for such an experiment, where all these issues are taken into account, is nevertheless beyond the scope of our present work,  that we hope will stimulate an experimental attempt.

%\section{Acknowledgments}
\begin{acknowledgments}
This work was supported by the Icelandic Research Fund, project 163438. 
\end{acknowledgments}

% data statement
%\section*{Data availability Statement}
%The data that support the findings of this study are available from the corresponding author upon reasonable request.

\bigskip

%\bibliographystyle{unsrt}
%\bibliographystyle{aip}
%\bibliography{Bibliography.bib}


%\appendix*
%\section{Supplementary Material}

\onecolumngrid

%\clearpage
%\newpage

%\appendix*
%\section{Supplementary Material}

\newpage 

\section*{Appendix}

\begin{figure}[h!]
\centering
%png version
%\includegraphics[scale=0.3]{1.png}
%\includegraphics[scale=0.3]{2.png}
%\includegraphics[scale=0.3]{3.png}
%\includegraphics[scale=0.3]{4.png}
%\includegraphics[scale=0.3]{5.png}
%\includegraphics[scale=0.3]{6.png}
%\includegraphics[scale=0.3]{7.png}
%\includegraphics[scale=0.3]{8.png}
%\includegraphics[scale=0.3]{9.png}
%\includegraphics[scale=0.3]{10.png}
%\includegraphics[scale=0.3]{11.png}
%\includegraphics[scale=0.3]{12.png}
%pdf version
\hspace{-50 mm}
\includegraphics[scale=1.0]{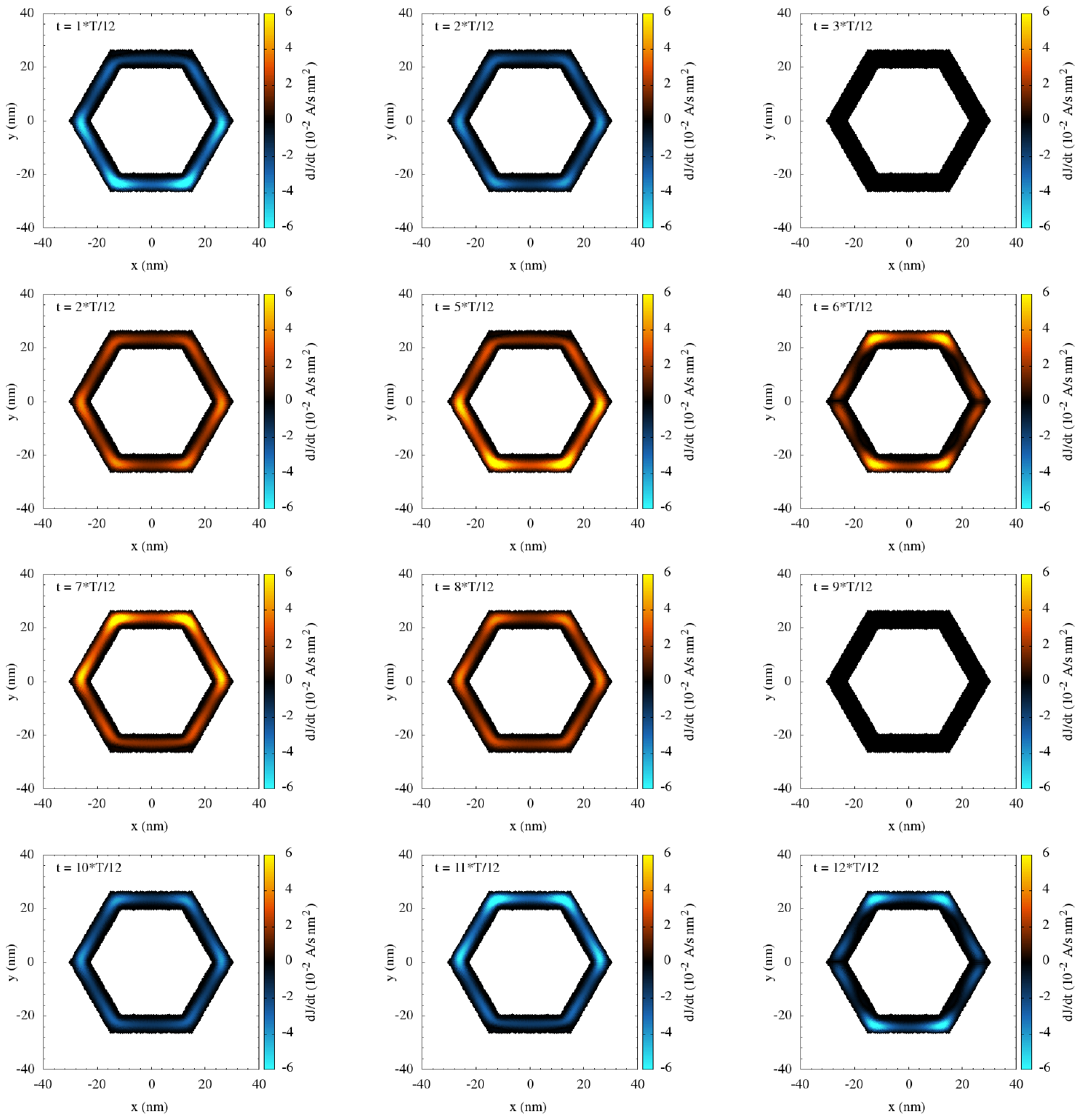}
\vspace{-100 mm}
\caption{Time-derivative of the current density distribution over a complete period when a magnetic field of $B = 1$ T is applied perpendicular to one of the edges of the nanowire. The carrier concentration is 
%$n = 10^{-5}$ $\mathrm{nm^{-3}}$.
$n = 10^{16}$ cm$^{-3}$
}
\label{Supplementary_a}
\end{figure}
\newpage

\begin{figure}[h!]
\centering
%png version
%\includegraphics[scale=0.3]{1a.png}
%\includegraphics[scale=0.3]{2a.png}
%\includegraphics[scale=0.3]{3a.png}
%\includegraphics[scale=0.3]{4a.png}
%\includegraphics[scale=0.3]{5a.png}
%\includegraphics[scale=0.3]{6a.png}
%\includegraphics[scale=0.3]{7a.png}
%\includegraphics[scale=0.3]{8a.png}
%\includegraphics[scale=0.3]{9a.png}
%\includegraphics[scale=0.3]{10a.png}
%\includegraphics[scale=0.3]{11a.png}
%\includegraphics[scale=0.3]{12a.png}
%pdf version
\hspace{-50 mm}
\includegraphics[scale=1.0]{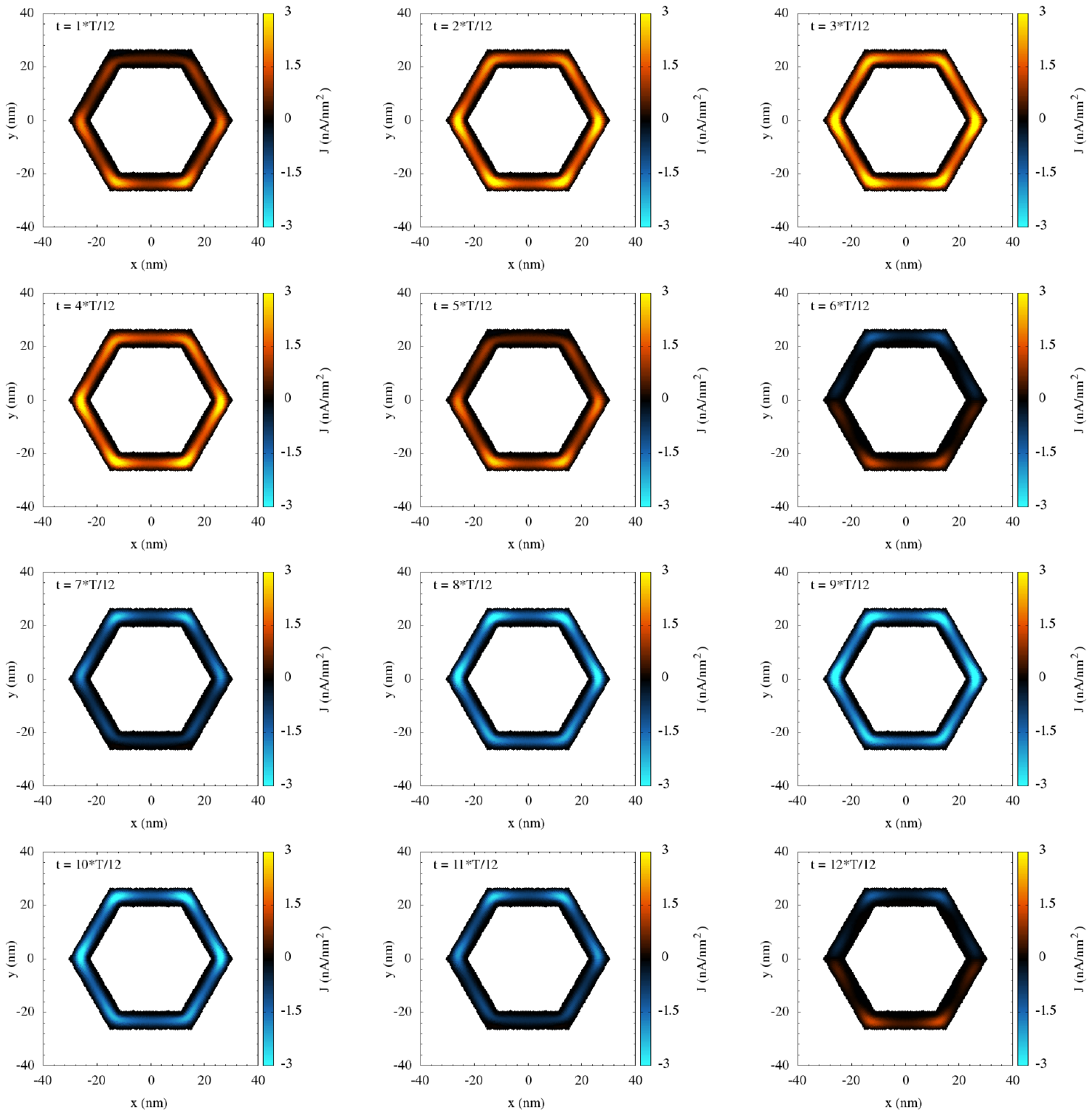}
\vspace{-100 mm}
\caption{Current density distribution over a complete period when a magnetic field of $B = 1$ T is applied perpendicular to one of the edges of the nanowire. The carrier concentration is 
%$n = 10^{-5}$ $\mathrm{nm^{-3}}$.
$n = 10^{16}$ cm$^{-3}$
}
\label{Supplementary_b}
\end{figure}
\newpage

\end{document}